\documentclass[english,prl,groupedaddress,reprint]{revtex4-1}
\usepackage[T1]{fontenc}
\usepackage[latin9]{inputenc}
\usepackage{mathrsfs}
\usepackage{amsmath}
\usepackage{amssymb}
\usepackage{graphicx}
\usepackage{braket}
\usepackage{xcolor}
\usepackage{comment}
\usepackage{bbold}
\usepackage[unicode=true,
 bookmarks=false,
 breaklinks=false,pdfborder={0 0 1},backref=false,colorlinks=true]
 {hyperref}
\hypersetup{
 linkcolor=blue,citecolor=blue,urlcolor=blue,linkcolor=blue,citecolor=blue,urlcolor=blue}
\makeatletter
\usepackage{microtype} 

\newcommand{\zr}{{\mathcal{Z}}_{\mathcal{R}}}
\newcommand{\zt}{{\mathcal{Z}}_{\mathcal{T}}}

\newcommand{\zrn}{{\tilde{\mathcal{Z}}_{\mathcal{R}}}}
\newcommand{\ztn}{{\tilde{\mathcal{Z}}_{\mathcal{T}}}}

\DeclareMathAlphabet{\mathscr}{OT1}{pzc}{m}{it} 
\newcommand*{\tr}[2][]{\ensuremath{\text{Tr}_{#1}\left[ #2 \right]}}

\newcommand{\id}{\ensuremath{{\mathbb{1}}}}

\begin{document}
\title{Many-body topological invariants from randomized measurements}
\author{Andreas Elben\textsuperscript{1,2}, Jinlong Yu\textsuperscript{1,2}, Guanyu Zhu\textsuperscript{3,a}, Mohammad Hafezi\textsuperscript{3,4}, Frank Pollmann\textsuperscript{5,6}, Peter Zoller\textsuperscript{1,2}, and Beno\^it Vermersch\textsuperscript{1,2,*}}

	\affiliation{\textsuperscript{1}{Center for Quantum Physics, University of Innsbruck, Innsbruck A-6020, Austria.}}
	\affiliation{\textsuperscript{2}{Institute for Quantum Optics and Quantum Information of the Austrian Academy of Sciences,  Innsbruck A-6020, Austria.}}
	\affiliation{\textsuperscript{3} {Joint Quantum Institute, NIST/University of Maryland, College Park, Maryland 20740 USA.}}
	\affiliation{\textsuperscript{4} {Institute for Research in Electronics and Applied Physics, University of Maryland, College Park, MD 20742, USA.}}
\affiliation{	\textsuperscript{5}{Department of Physics, Technical University of Munich, 85748 Garching, Germany.}}
\affiliation{	\textsuperscript{6}{Munich Center for Quantum Science and Technology (MCQST), Schellingstr. 4, D-80799 M\"unchen, Germany.}}
\affiliation{	\textsuperscript{a}{Current affiliation: IBM T.J. Watson Research Center, Yorktown Heights, New York 10598, USA.}}
\affiliation{	\textsuperscript{*}{Corresponding author.}}

\begin{abstract}
The classification of symmetry-protected topological (SPT) phases in one dimension has been recently achieved, and had a fundamental impact in our understanding of quantum phases in condensed matter physics.
In this framework, SPT phases can be identified by many-body topological invariants, which are quantized non-local correlators for the many-body wavefunction. 
While SPT phases can now be realized in interacting synthethic quantum systems, the direct  measurement of quantized many-body topological invariants has remained so far elusive.
Here, we propose measurement protocols for many-body topological invariants for all  types of protecting symmetries of one-dimensional interacting bosonic systems. 
Our approach relies on randomized measurements implemented with local random unitaries, and can be applied to any spin system with single-site addressability and readout. Our scheme thus provides a versatile toolbox to experimentally classify interacting SPT phases.
\end{abstract}
\maketitle

Symmetry-protected topological (SPT) phases are bulk-gapped phases with short-range entanglement: They are topologically non-trivial only in presence of certain symmetries but adiabatically connected to a trivial product state once these symmetries are explicitly broken~\cite{Senthil2015,Wen2017}.
The classification of SPT phases is based on  the  cohomology of the protecting symmetry groups~\cite{Wen2012-SPT-Classification}: Two quantum states belong to the same phase if the action of the symmetry group on the  states is realized by the same class of projective representations.
Many-body topological invariants (MBTIs) of unidimensional bosonic (or spin) SPT phases have been introduced to identify such projective representations directly from the many-body wave function~\cite{Cirac2012,Pollman-Turner2012,Shapourian-Ryu2017}.
MBTIs take a non-zero quantized value, for example $\pm 1$ for a  symmetry group with two  projective representations,  for any realization of a given SPT phase, and thus serve as a unique identifier of the SPT phase.  MBTIs can be therefore seen as generalizations of string order parameters that were introduced~\cite{denNijs1989} to detect certain SPT phases protected by internal symmetries. MBTIs can in particular identify SPT phases, in the absence of internal symmetries, and therefore of string order~\cite{Pollmann2010,Pollman-Turner2012}.

Recently,  the first experimental observations of topological phases in quantum simulators have been reported~\cite{Jotzu2014,Flaschner2016,Aidelsburger2015,DeLeseleuc2018}, offering new possibilities to probe, understand, and classify topological quantum matter.
On the single-particle level, topological phases have been detected by the measurement of the Chern number with ultra-cold atoms~\cite{Aidelsburger2015}, which can be regarded as the single-particle analog of MBTIs. An interacting SPT phase, i.e., the Haldane phase~\cite{HaldanePhase1983},  has been realized in a quantum simulator based on  Rydberg atoms~\cite{DeLeseleuc2018}. String orders have finally been measured in different scenarios~\cite{Endres2011,Hilker2017,DeLeseleuc2018}, and used in particular to reveal the presence of the Haldane phase~\cite{DeLeseleuc2018}.
In view of this experimental progress, a fundamental question emerges: How to classify different topological quantum phases experimentally, i.e., how to measure (highly non-local) 
MBTIs of interacting SPT phases?

In this letter, we show how to measure MBTIs in state-of-the-art experiments.
Our approach is based on randomized measurements, which consist in applying to a quantum state a sequence of random operations before performing measurements~\cite{VanEnk2012,Elben2018,Brydges2019,Vermersch2018a}.
Using local random unitaries, which can be realized in experiments with very high fidelity, and whose random distributions reflect different symmetry groups, 
we show that MBTIs can be directly extracted from the statistics of randomized measurements.

\begin{figure*}[tbh]
    \centering
    \includegraphics[width=0.95\textwidth]{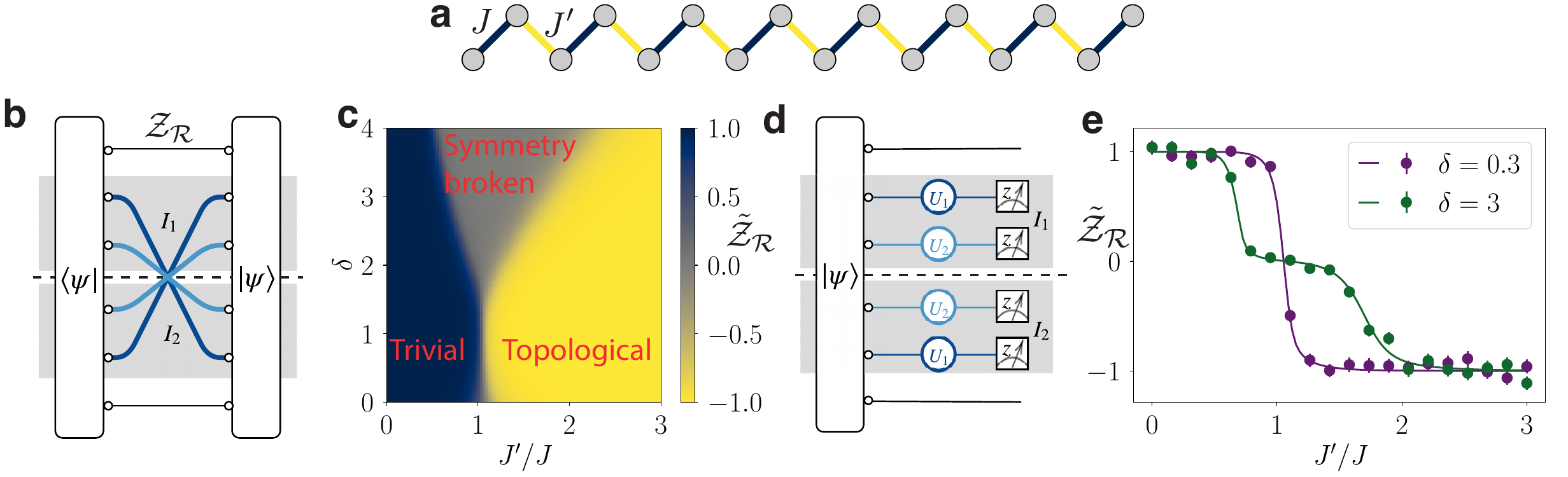}
    \caption{\textbf{Measuring the MBTI $\zr$ for the spin-$1/2$ bond-alternating XXZ model.}
    \textbf{a} Schematic of the model Eq.\eqref{Eq:Hami_XXZ}. 
    \textbf{b}, The partial reflection invariant $\zr$ [Eq.\eqref{Eq:Z_R}]  is defined as the expectation value of a partial reflection operator $\mathcal{R}_I$ (visualized by the blue lines).
    \textbf{c}, In terms of the normalized invariant $\zrn$, the full phase diagram of the bond-alternating XXZ model is revealed, here for a system size of $N=48$ spins and $n=6$ reflected pairs of spins. We find three phases with different quantized values of $\zrn$.
    \textbf{d}, Protocol to measure $\zrn$ via statistical correlations between randomized measurements, implemented with local random unitaries applied symmetrically around the central bond.
    \textbf{e}, The results of simulated experiments allow to identify topological phase transitions. The solid lines are results from DMRG, whereas the dots with error bars represent estimations from simulated randomized measurements with $N_U=512$ unitaries and $N_M=256$ measurements per unitary. The values agree within the  statistical errors that originate from a finite value of $N_U$ and $N_M$
    ; see text.}
    \label{fig:Figure1}
\end{figure*}
For concreteness, we present our approach in the context of the one-dimensional spin-$1/2$ bond-alternating XXZ model (Fig.~\ref{fig:Figure1}\textbf{a})
\begin{align}\label{Eq:Hami_XXZ}
H_{\text{XXZ}} &= \frac{J}{2}\sum\limits_{i=1}^{N/2} \left( {\sigma _{2i - 1}^ x \sigma _{2i}^ x + \sigma _{2i - 1}^ y \sigma _{2i}^ y }  + {\delta} \sigma^z_{2i-1}\sigma^z_{2i} \right) \nonumber \\
& + \frac{J^\prime}{2}\sum\limits_{i=1}^{N/2-1} \left( {\sigma _{2i}^ x \sigma _{2i + 1}^ x  + \sigma _{2i}^ y \sigma _{2i + 1}^ y}  + {\delta} \sigma^z_{2i}\sigma^z_{2i+1}  \right).
\end{align}
Here, $\sigma_i^\mu$ ($\mu=x,y,z$) are the Pauli matrices for the spin state at site $i$. $J$ and $J^\prime$ are alternating nearest-neighbor (NN) spin exchange coefficients, and $\delta$ denotes the exchange anisotropy. The case of $\delta = 1$ corresponds to the bond-alternating Heisenberg model, whereas the case of $\delta = 0$ corresponds to the bosonic version of the Su-Schrieffer-Heeger model~\cite{SSH1979} as realized recently with Rydberg atoms~\cite{DeLeseleuc2018}. Note that, expect for $\delta=0$ and $\delta=1$, the model is not integrable and thus has no single-particle correspondence. Our approach can be generalized to other spin Hamiltonians, e.g., spin-$1$ Haldane chain~\cite{HaldanePhase1983}, straightforwardly.

As shown below, the model \eqref{Eq:Hami_XXZ} hosts three different phases: a  trivial phase, a topological phase, and a symmetry-broken antiferromagnetic phase.
The trivial and topological phases are SPT phases protected by any one of the following three symmetries~\cite{Pollman-Protecting-Symmetries2012}: reflection (inversion) symmetry at the center bond, time-reversal symmetry, and dihedral group $D_2$ of  $\pi$-rotations of spins around the $x$, $y$, and $z$ axes. 

We now show how to measure MBTIs via random measurements.
First, SPT phases protected by reflection symmetry can be classified using the partial reflection MBTI $\zrn=\mathcal{Z}_\mathcal{R}/\sqrt{[\text{Tr}(\rho _{{I_1}}^2)+\text{Tr}(\rho _{{I_2}}^2)]/2 }$ \cite{Pollman-Turner2012}, with 
\begin{equation}\label{Eq:Z_R}
\zr ={\text{Tr}}\left[  \rho_I  \mathcal{R}_{I} \right].
\end{equation}
Here, $\rho_I=\mathrm{Tr}_{S-I}(\ket{\psi}\bra{\psi})$ is the reduced density matrix of the groundstate $\ket{\psi}$, and the interval $I=I_1 \cup I_2$ consists of two partitions $I_1$, $I_2$, each with $n$ sites. 
The non-local operator $\mathcal{R}_I$  ``spatially swaps'' $I_1$ and $I_2$ with respect to the reflection center. On 
every basis state $\ket{\mathbf{s}_{I}}=\ket{s_1,s_2,\dots,s_{2n}}$ ($s_i=\uparrow,\downarrow$ for $i \in I$), 
it acts as: $\mathcal{R}_I\ket{\mathbf{s}_{I}}=\ket{ s_{2n},s_{2n-1},\dots,s_1}\equiv \ket{\mathcal{R}_I(\mathbf{s}_{I})}$. This operation is graphically  shown  in Fig.~\ref{fig:Figure1}\textbf{b}, where the state of each site of $I$, represented as a blue line, is ``contracted'' with the state of the mirror symmetric site.

The MBTI $\zrn$ probes the action of the reflection symmetry on the many-body  state $\ket{\psi}$. Using tensor-network theory, one can show analytically that, for a gapped many-body Hamiltonian (e.g. $H_\text{XXZ}$), $\zrn$ approaches a quantized value in the thermodynamic limit $n,N\to \infty$~\cite{Pollman-Turner2012}.  The typical value of $n$ required to achieve convergence is determined by the correlation length in the system and is discussed in detail below.  For our model Eq.~\eqref{Eq:Hami_XXZ}, the phase-diagram evaluated by the MBTI $\zrn$, calculated numerically using the density-matrix-renormalization-group (DMRG) method, is shown in Fig.~\ref{fig:Figure1}\textbf{c}, where
the three phases can be identified: (i) A phase with anti-ferromagnetic order where reflection is spontaneously broken with $\zrn = 0$; (ii) The trivial SPT 
phase with $\zrn = +1$; (iii) The 
nontrivial
SPT phase with $\zrn = -1$.

The MBTI $\zrn$, which is a highly non-local and non-linear functional of the reduced density matrix $\rho_I$ can be measured with randomized measurements, with the following recipe:
(i) One first prepares the groundstate $\ket{\psi}$ via, e.g., adiabatic state preparation (see Supplementary Information 
for details). 
(ii.a) One applies to $\ket{\psi}$ a set of local random unitaries whose spatial distribution is reflection symmetric. This corresponds to a unitary operation $U_\mathcal{R}$ of the form $
U_\mathcal{R}=\bigotimes_{i=1}^{2n} U_i$, with $U_i =U_{2n-i+1}$ for $i=1,\dots, n$. The unitaries $U_i$ are drawn randomly from the circular unitary ensemble (CUE) defined on the local Hilbert space of spin $i$. Such unitaries can be generated with high fidelity in quantum simulators with single-site control, as shown in recent experiments~\cite{Brydges2019}.
(ii.b) One measures the occupation probabilities $P_{U_\mathcal{R}}(\mathbf{s}_{I})=\bra{\mathbf{s}_{I}} U_\mathcal{R} \rho_I U^\dagger_\mathcal{R} \ket{\mathbf{s}_{I}}$ of the basis states $\mathbf{s}$, by performing projective measurements in the basis $\mathbf{s}$.
(iii) One repeats (i)-(ii)  for  many independently sampled random unitaries  $U_\mathcal{R}$.

Given the set of outcome probabilities $P_{U_\mathcal{R}}(\mathbf{s}_{I})$, one obtains first $\zr$ from
\begin{align}
    \zr = 2^{n} \sum_{\mathbf{s}_{I}}  (-2)^{-\frac{1}{2} D[\mathbf{s}_{I}, \mathcal{R}_I(\mathbf{s}_{I})] } \; \overline{P_{U_\mathcal{R}}(\mathbf{s}_{I})} \; .
    \label{eq:zr_prob}
\end{align}
Here, $\overline{\vphantom{h} \cdots }$ denotes the ensemble average over the random unitaries and  $D[\mathbf{s}_{I}, \mathcal{R}(\mathbf{s}_{I})]\equiv \#\left\{i \in I| s_i\neq s_{2n-i+1}\right\}$
is
the  Hamming distance between $\ket{\mathbf{s}_I}$ and $\ket{\mathcal{R}_I(\mathbf{s}_{I})}$.  Equation~\eqref{eq:zr_prob} can be proven using the $2$-design identities of the CUE (see Methods) and shows that the MBTI $\zr$ can be directly extracted from the statistics of randomized measurements.  Secondly, the purity  $\text{Tr}(\rho _{{I_1}}^2)$ (and similarly $\text{Tr}[\rho _{{I_2}}^2]$) is estimated using the relation~\cite{Elben2018,Brydges2019}
\begin{align}
\text{Tr}(\rho _{{I_1}}^2) = 2^{n} \sum_{\mathbf{s}_{I_1},\mathbf{s'}_{I_1}}  (-2)^{- D[\mathbf{s}_{I_1},\mathbf{s'}_{I_1}] } \; \overline{P_{U_\mathcal{R}}(\mathbf{s}_{I_1})P_{U_\mathcal{R}}(\mathbf{s'}_{I_1})},
\label{eq:purity_prob}
\end{align}
with the reduced probabilities $P_{U_\mathcal{R}}(\mathbf{s}_{I_1})=\mathrm{Tr}(\ket{\mathbf{s}_{I_1}}\bra{\mathbf{s}_{I_1}}U_\mathcal{R} \rho_I U^\dagger_\mathcal{R})$.
Thus, we obtain the normalized MBTI from the second-order correlations of randomized measurements, implemented with local random operations with a  distribution that is tailored to identify a certain symmetry (here, the reflection symmetry) of the quantum state.  This is the key idea in our approach and we show below how to apply it to measure any MBTI.
For illustration, we show in Fig.~\ref{fig:Figure1}\textbf{e} the value of $\zrn$, (i) calculated from the DMRG method (line), and (ii) estimated from simulated randomized measurements (dots).

\begin{figure*}[hbt]
    \centering
    \includegraphics[width=0.75\textwidth]{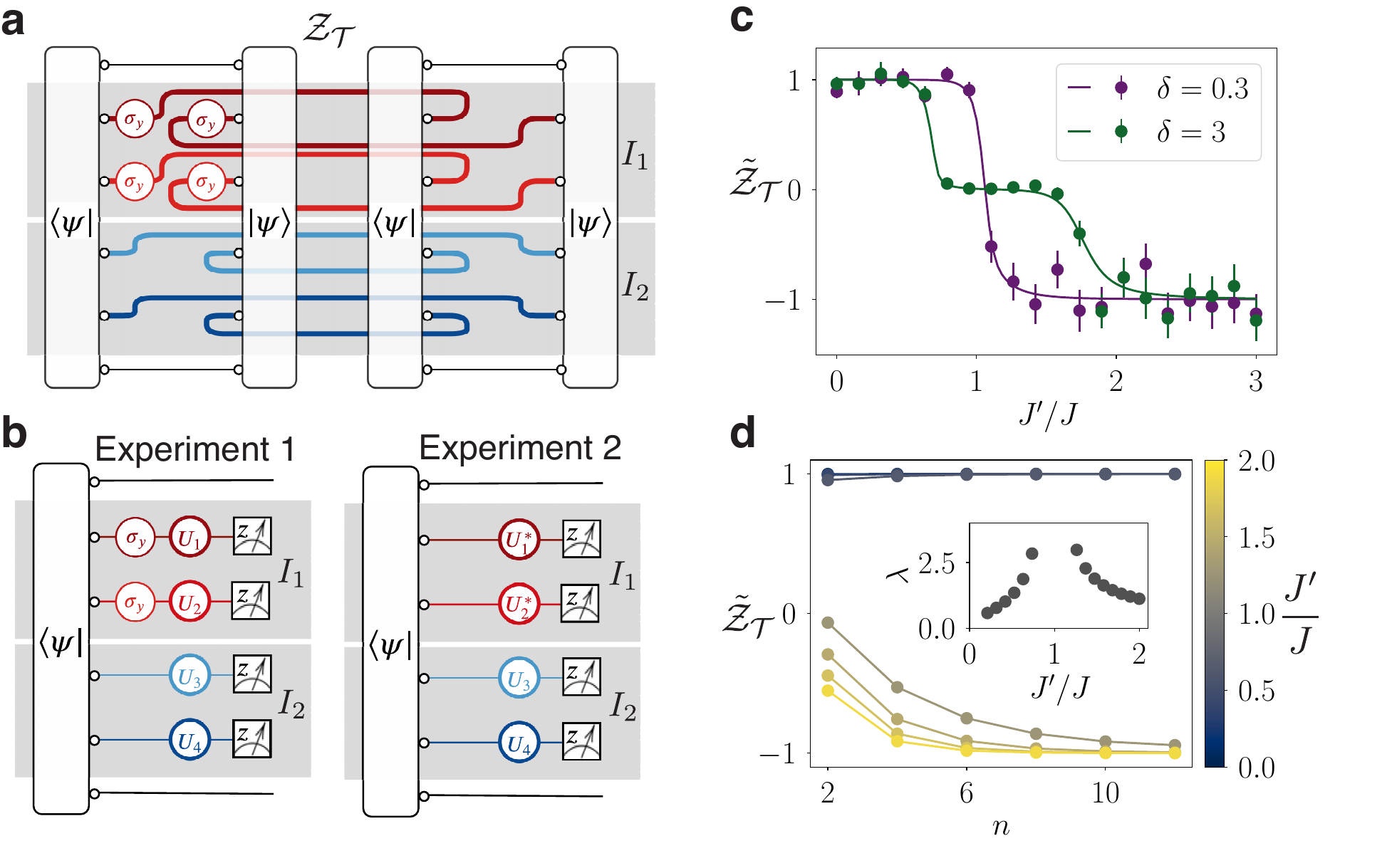}
    \caption{\textbf{Measurement of the MBTI $\zt$ with randomized measurements.}
    \textbf{a}, Graphical representation of the definition of the time-reversal invariant $\mathcal{Z}_\mathcal{T}$ [Eq.~\eqref{Eq:Z_T}] involving partial transpose (red lines) and partial SWAP (blues lines) operations.
    \textbf{b}, Experimental protocol to measure $\mathcal{Z}_\mathcal{T}$ with two experiments which are correlated using randomized measurements. To account for the anti-unitarity of the time reversal symmetry, the local random unitaries  applied in $I_1$ (red) in the two experiments  are complex conjugate to each other.
    \textbf{c}, Simulated measurements of $\zrn$ (dots with statistical error bars, with $N_U=768$, $N_M=512$) revealing the topological phase transitions in the bond-alternating XXZ-model as a function of $J'/J$ for two values of $\delta$. Solid lines are calculated with the DMRG method, in a system with $N=48$ sites, and $n=6$ per interval $I_1$ and $I_2$.
    \textbf{d}, $\tilde{\mathcal{Z}}_\mathcal{T}$ converges as a function of the partition size $n$ to the quantized values $\pm 1$. Different colors represent different values of $J'/J$. Inset: The divergence of the corresponding correlation length $\lambda$, extracted from an exponential fit on the first three values of $n$, can be used to detect the quantum critical point.}
    \label{fig:Figure2}
\end{figure*}

 We now present the protocol to measure the MBTI associated with the time-reversal symmetry~\cite{Pollman-Turner2012}  $\ztn=\mathcal{Z}_\mathcal{T}/([\text{Tr}(\rho _{{I_1}}^2)+\text{Tr}(\rho _{{I_2}}^2)]/2)^{3/2}$, with
\begin{equation}\label{Eq:Z_T}
   \zt = {\text{Tr}}\left[ {{\rho _I}{u_{\mathcal{T}}}\rho _I^{{{T}_1}}u_{\mathcal{T}}^\dag } \right].
 \end{equation} 
Here, ${T}_1$ denotes the partial transpose operation on the partition $I_1$, and $u_\mathcal{T}=\prod_{i\in I_1} \sigma_i^y$. The contraction operation resulting in $\zt$ is illustrated graphically in Fig.~\ref{fig:Figure2}\textbf{a}.

The MBTI $\ztn$ is a non-linear functional of two copies of the (partially transposed) density matrix $\rho_I$, which can be measured via the following recipe (Fig.~\ref{fig:Figure2}{\bf b}).
After (i) the state preparation, we perform two experiments:  
(ii.a.1) In the first experiment, we apply $U_\mathcal{T}^{(1)}=U_{I_1}u_\mathcal{T} \otimes U_{I_2}$, with $U_{I_1} =\bigotimes_{i=1}^{n} U_i$ and $U_{I_2} =\bigotimes_{i=n+1}^{2n} U_i$, each $U_i$ being taken independently from the CUE. 
(ii.b.1) We measure the probabilities $P_{U^{(1)}_\mathcal{T}}(\mathbf{s}_{I})$, see the left panel of  Fig.~\ref{fig:Figure2}\textbf{b}.
(ii.a.2) In a second experiment, we use the unitaries $U_\mathcal{T}^{(2)}=U^{*}_{I_1} \otimes U_{I_2}$.
(ii.b.2) We measure $P_{U^{(2)}_\mathcal{T}}(\mathbf{s}_{I})$, see the right panel of Fig.~\ref{fig:Figure2}\textbf{b}.
(iii) We repeat the two experiments (i-ii) with different unitaries $U_i$ and estimate
\begin{align}
    \zt = 2^{2 n} \sum_{\mathbf{s}_{I},\mathbf{s}^\prime_{I}}  (-2)^{-D[\mathbf{s}_I,\mathbf{s}^\prime_I]} \overline{P_{U^{(1)}_\mathcal{T}}(\mathbf{s}_{I})P_{U^{(2)}_\mathcal{T}}(\mathbf{s}^\prime_{I})}\label{eq:zt_prob}
\end{align}
from cross-correlations of the two experiments. In addition, the purity to normalize $\zt$ to $\ztn$ is obtained from the same experimental data using the relation Eq.~\eqref{eq:purity_prob}.

Equation~\eqref{eq:zt_prob}, which is also proven in the Methods, shows that the partial time-reversal MBTI can be accessed from correlations between measurements using random unitary operations which are complex-conjugated. In Fig.~\ref{fig:Figure2}\textbf{c}, we compare  values of $\ztn$ obtained with the DMRG method with the ones estimated from finite number of randomized measurements. In Fig.~\ref{fig:Figure2}\textbf{d}, we also show that by extracting $\ztn$ (or $\zrn$) for different $n$, one can measure the correlation length $\lambda$ of SPT phases, i.e., the characteristic length above which MBTIs become quantized. In particular, one can identify quantum critical points separating different SPT phases from the divergence of $\lambda$.

The two examples given above illustrate how to access MBTIs from the statistics of measurements performed after correlated local random unitary operations. In the Supplementary Information, we show how to access  MBTIs for internal symmetries and combination of symmetries, and also how to identify the breaking/protection of different symmetries. This provides a complete set of protocols to experimentally probe the classification of one-dimensional bosonic SPT phases.

We now comment on various potential sources of errors in implementing our protocol.
First, statistical errors are due to the finite number of repetitions of the experiment used to estimate the statistical correlations between randomized measurements. 
As detailed in the Supplementary Information, we find that the typical required number of measurements to access MBTIs within a given accuracy (scaling as $2^{1.5n}$ to access $\zr$ for instance) are very similar to the requirements to measure state purities~\cite{Elben2018,Brydges2019}, and thus compatible with state-of-the-art experimental platforms of Rydberg atoms, trapped ions, and superconducting qubits with high repetition rates.
Randomized measurements also feature 
a natural robustness with respect to decoherence, readout errors, errors in the implementations of random unitaries~\cite{Vermersch2018a}, since they are based on extracting relevant quantities from ensemble averages (and not from individual measurements). We expect thus our protocols to allow faithful measurements of MBTIs in various experimental platforms.

Finally, in a quantum simulation experiment, the ground state of a spin model is typically prepared via adiabatic-state-preparation~\cite{DeLeseleuc2018}.
The MBTIs $\zrn(t)$ and $\ztn(t)$, defined with respect to the time-evolved wavefunction  $\ket{\Psi(t)}$, can be used to verify the preparation of an SPT phase, and measure the corresponding correlation length, c.f., Supplementary Information. Similarly, our protocols can  be used to probe topology in non-equilibrium systems~\cite{Cooper2018}.

To conclude, the use of randomized measurements to measure topological properties of the many-body wavefunction is a new paradigm that enables the experimental classification of one-dimensional SPT phases.
Our work also opens the possibilities for probing two-dimensional SPT phases~\cite{Zaletel2014}, but also quantum phases with an intrinsic topological order, where long-range entanglement is the mechanism behind topological protection~\cite{Wen1990}.
In particular, modular matrices revealing anyonic statistics~\cite{Wen1990} can be expressed as spatial reflection operators in a form analog to $\zrn$ on torus geometries~\cite{modular_transformation2017}, and could thus be measured via random measurements, complementing interferometric approaches based on impurities~\cite{Grusdt2016}.
Once the intrinsic topological order has been identified in terms of anyonic statistics, an even finer classification can be achieved when taking into account topological symmetries, i.e., classifying different symmetry-enriched topological (SET) orders~\cite{Essin2013, Huang2014,barkeshliReflection2016,Garre2019}.  These symmetries, like reflection and time-reversal \cite{barkeshliReflection2016}, can be distinguished via the same MBTIs as for SPT phases (defined in compactified 1D geometries), and thus could also be probed via random measurements.

{\bf Acknowledgments}\\
We thank A. Browaeys, Z.P Cian, I. Cirac,  M. Hermele, M. Knap, T. Lahaye, Z.-X. Liu, and S. Montangero for discussions.
The tensor-network simulations (DMRG and TEBD) were realized using the ITensor Library (http://itensor.org). 
Research in Innsbruck is supported by the ERC Synergy Grant UQUAM, the project PASQUANS of the EU Quantum Technology flagship, and the Simons foundation via the Simons collaboration UQM.
Research in Maryland was supported by ARO-MURI and NSF-PFC at the JQI.
FP acknowledges funding from the Deutsche Forschungsgemeinschaft (DFG, German Research Foundation) under Germany's Excellence Strategy - EXC-2111-390814868,  DFG TRR80, Project number 107745057, 
DFG Research Unit FOR 1807 through grant no. PO 1370/2- 1,  
DFG TRR80, Project number 107745057, and the 
European Research Council (ERC) under the European Unions Horizon 2020 research, and innovation program grant agreement no. 771537.

A.E. and J.Y. contributed equally to this work. 

\bibliographystyle{apsrev4-1}
%

\appendix
\section{Topological invariants from randomized measurements}
In this section, we present the proofs of 
Eqs.~\eqref{eq:zr_prob}-\eqref{eq:zt_prob}, relating MBTIs to statistical correlations of randomized measurements.  
As in the main text, we focus on the case of spin-$1/2$ systems. Our formulas can, however, be extended straightforwardly to the cases with higher internal dimensions (spins $1$, $3/2$, \dots).

\subsection{Random unitary calculus}
We begin by summarizing elementary properties of random unitaries from the circular unitary ensemble (CUE).
We discuss the minimal case of two spins, each with Hilbert space $\mathcal{H}$. These can be either (i) two spins located at different lattice sites in a single many-body system (partial inversion invariant) or (ii) two spins located at the same site but realized in two different, sequentially performed, experiments (time reversal invariant).
Given a two-spin operator $O$ acting on both spins with total Hilbert space $\mathcal{H}^{\otimes 2}$, we define the unitary twirling channel 
\begin{align}
\Phi(O) \equiv \overline{U^\dagger \otimes U^\dagger O U\otimes U},
\end{align}
where $\overline{\vphantom{h}\dots}$ denotes the average over random unitaries $U$ taken from the CUE (i.e., the average with respect to the Haar measure on the group of unitary matrices on $\mathcal{H}$). Using the $2$-design identities of the CUE, we find~\cite{Elben2018a}
\begin{align*}
    \Phi(O) &= \frac{1}{3} \left(\tr[]{O}  - \frac{1}{2} \tr[]{ \mathbb{S}O} \right) \id_2  \nonumber \\ & +  \frac{1}{3} \left(\tr[]{ \mathbb{S} O}  - \frac{1}{2} \tr[]{O} \right)  \mathbb{S},
\end{align*}
where $\mathbb{S}= \sum_{s,s'}\ket{s,s'}\bra{s',s}$ denotes the swap operator.
We also define the closely related isotropic twirling channel~\cite{Watrous2018} \begin{align}
\Psi(O) \equiv \overline{U^\dagger \otimes (U^*)^\dagger O U\otimes U^*}  = \left[  \Phi  \left(O^{T_2}\right) \right]^{T_2}.
\end{align}
Here, $(\cdot)^{T_2}$ denotes the partial transpose with respect to the second spin.
For the following proofs, we will use an operator $  \tilde{O}\equiv 2 \sum_{s,s'} (-2)^{-D[s,s']}  \ket{s,s'}\bra{s,s'} $, which is diagonal in the computational basis, and fullfills~\cite{Elben2018a}
\begin{align}
        \Phi\left(\tilde{O}\right)&= \mathbb{S}, \label{eq:swap}\\
        \Psi\left(\tilde{O}\right)&= \mathbb{S}^{T_2}= \sum_{s,s'}\ket{s,s}\bra{s',s'}\equiv \mathbb{T} \label{eq:transpose}.
\end{align} In the following, we show how to use the identities \eqref{eq:swap} and \eqref{eq:transpose} to proof Eqs.~\eqref{eq:zr_prob}-\eqref{eq:zt_prob} relating randomized measurements and MTBIs.

\subsection{Partial reflection invariant}
The MBTI $Z_\mathcal{R}$ is inferred from statistical correlations of randomized measurements, performed on a quantum state $\rho_I$, which are implemented by applying spatially correlated local random unitaries of the form $
U_\mathcal{R}=\bigotimes_{i=1}^{2n} U_i$, with $U_i =U_{2n-i+1}$ for $i=1,\dots, n$.
To prove Eq.~\eqref{eq:zr_prob}, we first note that its right-hand side can be rewritten as an expectation value of an operator $O_\mathcal{R}$:
 \begin{eqnarray*}
\mathcal{E}_\mathcal{R} &\equiv& 2^{n} \sum_{\mathbf{s}_{I}}  (-2)^{ -\frac{1}{2} D[\mathbf{s}_{I}, \mathcal{R}_I(\mathbf{s}_{I})]} \overline{P_{U_\mathcal{R}}(s_I)}  \nonumber\\
 &=& \tr{ \overline{U_\mathcal{R}^\dag O_\mathcal{R} U_\mathcal{R}}\rho_I} \nonumber \\
 &=& \tr{ \bigotimes_{i=1}^{n} \overline{(U_i^\dag\otimes U_{i}^\dag) O_{\mathcal{R},i} (U_i \otimes U_{i}}) \rho_I}, 
 \end{eqnarray*}
 with $O_\mathcal{R}=\bigotimes_{i=1}^n O_{\mathcal{R},i}$, which is a tensor product of operators
\begin{eqnarray}
 O_{\mathcal{R},i} &=&  2 \sum_{\mathbf{s}_{I[i]}}  (-2)^{ -\frac{1}{2} D[\mathbf{s}_{I[i]}, \mathcal{R}_I(\mathbf{s}_{I[i]})]} \ket{\mathbf{s}_{I[i]}}\bra{\mathbf{s}_{I[i]}}
 \label{eq:Oi}
 \end{eqnarray}
 acting on pairs of spins $I[i]=(i,2n-i+1)$. We also used the independence of the unitaries  $U_i$ and  $U_{i'}$ ($i\neq i', i,i'=1,\dots, n$)  applied to different pairs of spins  $I[i]$ and $I[i']$, respectively. 
Using Eq.~\eqref{eq:swap} with the identification $\mathcal{R}_{I[i]} \to \mathbb{S}$, and $O_{\mathcal{R},i}\to \tilde O$, we find
 \begin{eqnarray}
 \overline{(U_i^\dag\otimes U_{i}^\dag) O_{\mathcal{R},i} (U_i \otimes U_{i}}) \nonumber 
  &=&  \mathcal{R}_{I[i]} 
 \end{eqnarray}
and therefore obtain
\begin{eqnarray*}
\mathcal{E}_\mathcal{R}=\tr{\bigotimes_{i=1}^n \mathcal{R}_{I[i]} \rho_I}=\zr.
\end{eqnarray*}

\subsection{Partial time-reversal invariant}
The MTBI $Z_\mathcal{T}$ is inferred from the statistical correlations of  correlated randomized measurements on two (sequential) experiments, both preparing a quantum state $\rho_I$. These are implemented by applying to the sites in an interval  $I=I_1\cup I_2$ local random unitaries $U_\mathcal{T}^{(1)}=U_{I_1}u_\mathcal{T} \otimes U_{I_2}$ (experiment 1) and $U_\mathcal{T}^{(2)}=U^*_{I_1} \otimes U_{I_2}$ (experiment 2) with $U_{I_{1,2}}=\bigotimes_{i \in I_{1,2}} U_i$ and $u_\mathcal{T}=\bigotimes_{i\in I_1} \sigma^y_i$, respectively (see main text). To prove Eq.~\eqref{eq:zt_prob}, we rewrite its right-hand side as 
\begin{align}
 \mathcal{E}_\mathcal{T} &\equiv   2^{2 n} \sum_{\mathbf{s}_{I},\mathbf{s}^\prime_{I}}  (-2)^{-D[\mathbf{s}_I,\mathbf{s}^\prime_I]} \overline{P_{U^{(1)}_\mathcal{T}}(\mathbf{s}_{I})P_{U^{(2)}_\mathcal{T}}(\mathbf{s}^\prime_{I})}
    \nonumber \\
        &= \tr{\overline{ (U_\mathcal{T}^{(1)})^\dag \otimes ( U_\mathcal{T}^{(2)})^\dag\,   O_\mathcal{T} \,U_\mathcal{T}^{(1)}\otimes U_\mathcal{T}^{(2)}} \,(\rho_I  \otimes \rho_I) }\nonumber \\
    &= \textrm{Tr}\left[  \bigotimes_{i\in I_1} \overline{ U_i^\dag \otimes (U_i^*)^\dag\,   O_{\mathcal{T},i} \,U_i\otimes U_i^*} \right. \nonumber \\
    &\qquad \quad \left. \bigotimes_{i\in I_2} \overline{ U_i^\dag \otimes U_i^\dag\,   O_{\mathcal{T},i} \,U_i\otimes U_i} \; (\tilde \rho_I  \otimes \rho_I) \right].
    \label{eq:ztprob}
\end{align}
Here, we defined $\tilde \rho_I\equiv (u_\mathcal{T} \otimes \mathbb{1}_{I_2}) \,  \rho_I  \, (u^\dagger_\mathcal{T} \otimes \mathbb{1}_{I_2})$ and  used the (spatial) tensor product structure of the operator $ O_\mathcal{T}= \bigotimes_{i \in I} O_{\mathcal{T},i}$  with
\begin{align}
     O_{\mathcal{T},i} &=  2 \sum_{s_i,s_i'} (-2)^{-D[s_{i},s_{i'}]}
 \ket{s_i}\bra{s_i} \otimes  \ket{s'_i}\bra{s'_i}.
 \label{eq:Oi_t}
\end{align}
Using Eqs.~\eqref{eq:swap} and \eqref{eq:transpose} with the identification $\mathbb{S}_i \to \mathbb{S}$, $\mathbb{T}_i \to \mathbb{T}$, and $O_{\mathcal{T},i}\to \tilde O$ we thus directly obtain 
\begin{align}
    \mathcal{E}_\mathcal{T}  &= \tr[]{\bigotimes_{i\in I_1} \mathbb{T}_i \bigotimes_{i\in I_2} \mathbb{S}_i \; (\tilde \rho_I  \otimes \rho_I) } \nonumber \\
    &=  \tr[]{ (\tilde  \rho_I)^{T_{I_1}}  \rho_I }  = \mathcal{Z_T}.
\end{align}

 \section{Internal symmetries and combinations of symmetries}
 In this section, we present randomized measurement protocols to access MBTIs associated with internal symmetries and combinations of symmetries.
 
 \subsection{Internal symmetries}
  \begin{figure*}[t]
    \centering
    \includegraphics[width=0.65\textwidth]{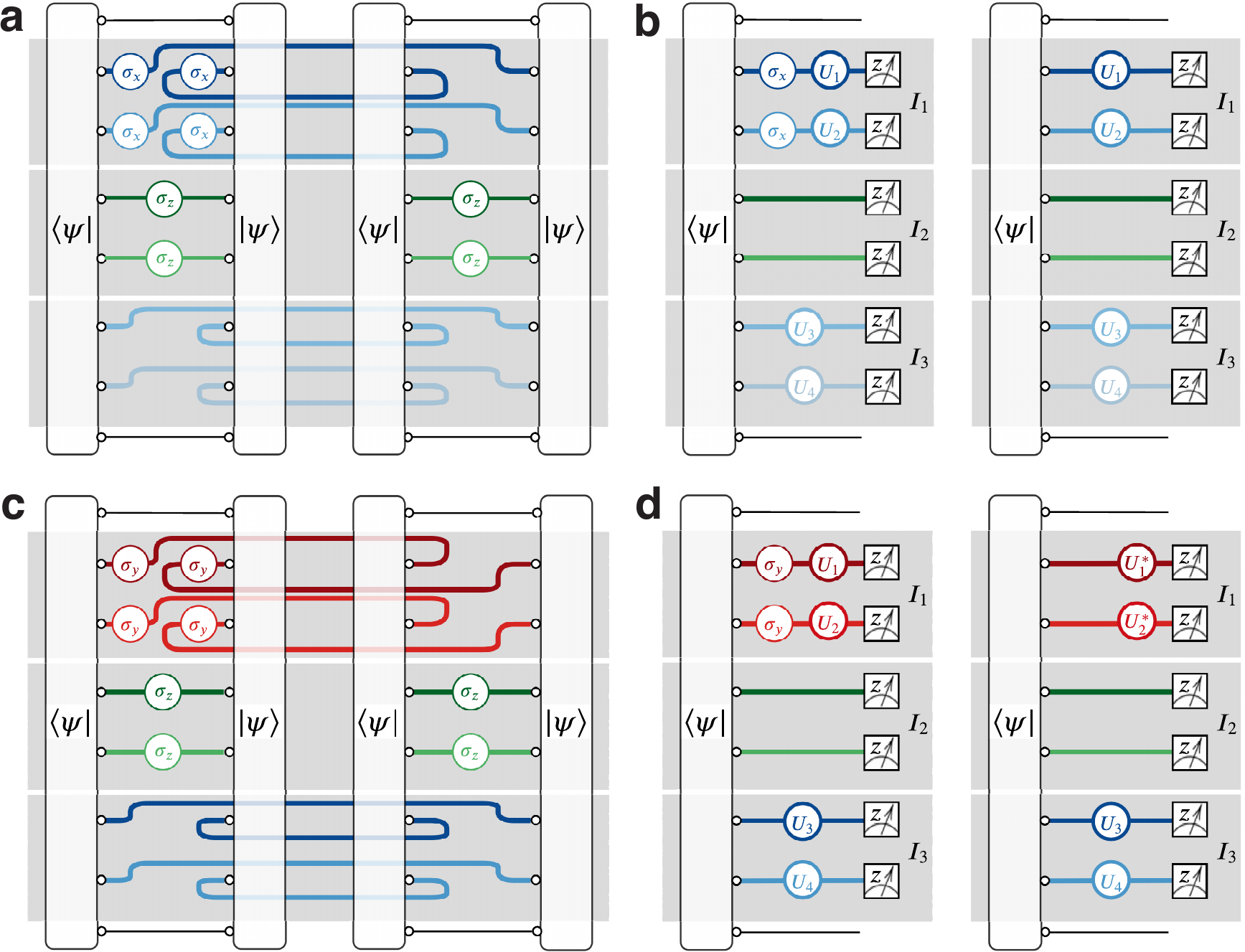}
    \caption{\textbf{Invariants for on-site symmetries and the combination of on-site and time-reversal symmetry.}
    {\bf a}, Invariant $Z_{D_2}$ for the $D_2$ internal symmetry and {\bf b}, protocol via randomized measurements.
    {\bf c} Klein-Bottle Invariant $Z_\mathrm{KB}$, and 
    {\bf d}, protocol.
    }
    \label{fig:invariantsSM}
\end{figure*}
 We first consider the $D_2$ invariant introduced in Ref.~\cite{Cirac2012}, which is in particular relevant for our model:
 \begin{equation}
 \mathcal{Z}_{D_2} = \tr{\mathbb{S}_{I_1}\mathbb{Z}_{I_2} \mathbb{S}_{I_3}   (\breve{\rho}_I \otimes \rho_I)}, 
 \end{equation}
 with $\mathbb{S}_{I_x}=\bigotimes_{i\in I_x} \mathbb{S}_i$, $\mathbb{Z}_{I_2}=(\bigotimes_{i\in I_2}\sigma_i^z) \otimes (\bigotimes_{i\in I_2}\sigma_i^z)$, 
 and $\breve{\rho}_I=(\bigotimes_{i\in I_1} \sigma_i^x) \rho_I (\bigotimes_{i\in I_1} \sigma_i^x)$.
 The MBTI $\mathcal{Z}_{D_2}$, represented graphically in Fig.~\ref{fig:invariantsSM}\textbf{a},  can be measured correlating two experiments using two sets of random unitaries,
 \begin{eqnarray}
 U^{(1)}_{D_2}&=&(\otimes_{i \in I_1} U_i \sigma_i^x)(\otimes_{i\in I_2} 1_i)(\otimes_{i \in I_3} U_i)
 \nonumber \\
  U^{(2)}_{D_2}&=&(\otimes_{i \in I_1} U_i)(\otimes_{i \in I_2} 1_i)(\otimes_{i \in I_3} U_i\nonumber),
  \end{eqnarray}
 as illustrated in Fig.~\ref{fig:invariantsSM}\textbf{b}. Here, $1_i$ is the identity matrix on spin $i$.
From a measurement in the computational basis, one can access 
 \begin{align}
 \mathcal{E}_{D_2} &= 2^{2n} \sum_{\mathbf{s}_{I},\mathbf{s}_{I}'}  (-2)^{- D[\mathbf{s}_{\{I_1,I_3\}},{s}_{\{I_1,I_3\}}')]}
 \sigma_{I_2}^z(s_{I_2}) \sigma_{I_2}^z(s'_{I_2})
  \nonumber \\
 & \hspace{1cm} \overline{P_{U^{(1)}_{D_2}}(s_I)P_{U^{(2)}_{D_2}}(s'_I)},
  \nonumber \\
  &= \tr{\overline{ (U_{D_2}^{(1)})^\dag \otimes  (U_{D_2}^{(2)})^\dag\,   O_{D_2} \,U_{D_2}^{(1)}\otimes U_{D_2}^{(2)}} \,(\rho_I  \otimes \rho_I)}\nonumber \\
    &= \textrm{Tr}\left[  \bigotimes_{i\in I_1} \overline{ U_i^\dag \otimes U_i^\dag\,   O_{D_2,i} \,U_i\otimes U_i} 
    \bigotimes_{i\in I_2} O_{D_2,i}
    \right. \nonumber \\
    &\qquad \quad \left. \bigotimes_{i\in I_3} \overline{ U_i^\dag \otimes U_i^\dag\,   O_{D_2,i} \,U_i\otimes U_i} \; (\breve{\rho}_I  \otimes \rho_I) \right].
 \end{align}
with $\sigma_{I_2}^z(s_{I_2})= \prod_{i\in I_2} \bra{s_i} \sigma_i^z \ket{s_i}$. Here, we have introduced the operator $O_{D_2}=\bigotimes_{i\in I}O_{D_2,i}$, with 
\begin{align*}
O_{D_2,i\in I_1 \cup U_3} &= 2 \sum_{s_i,s_i'} (-2)^{-D[s_{i},s_{i'}]}
 \ket{s_i}\bra{s_i} \otimes  \ket{s'_i}\bra{s'_i}. \\
O_{D_2,i\in I_2}&=\sigma_i^z \otimes \sigma_i^z.
\end{align*}
Using the random unitary twirling channel introduced in the Methods, this leads to
 \begin{align}
 \mathcal{E}_{D_2} &= \tr{ \bigotimes_{i\in I_1} \mathbb{S}_{i} \bigotimes_{i\in I_2} O_{D_2,i} \bigotimes_{i\in I_3} \mathbb{S}_{i} (\breve{\rho}_I  \otimes \rho_I)}
 \nonumber \\ 
 &= \mathcal{Z}_{D_2}.
 \end{align}

 \subsection{Combination of symmetries}
  Finally, with randomized measurements, we can access MBTIs associated with combinations of symmetries. 
 We consider, for instance, the Klein-Bottle invariant~\cite{Shiozaki2017}, associated with the rotation along one axis (here $z$) and time-reversal symmetry
  \begin{equation}
 \mathcal{Z}_\mathrm{KB} = \tr{\mathbb{S}_{I_1}\mathbb{Z}_{I_2} \mathbb{S}_{I_3}   (u_\mathcal{T} \rho_I^{T_1} u_\mathcal{T}^\dag \otimes \rho_I) }, 
 \end{equation}
 cf., Fig.~\ref{fig:invariantsSM}\textbf{c}. As shown in Fig.~\ref{fig:invariantsSM}\textbf{d}, the protocol to measure $\mathcal{Z}_\mathrm{KB}$ consists in applying two sets of unitaries:
 \begin{eqnarray}
 U^{(1)}_\mathrm{KB}&=&(\otimes_{i \in I_1} U_i \sigma_i^y)(\otimes_{i\in I_2} 1_i)(\otimes_{i \in I_3} U_i)
 \nonumber \\
  U^{(2)}_\mathrm{KB}&=&(\otimes_{i \in I_1} U_i^*)(\otimes_{i \in I_2} 1_i)(\otimes_{i \in I_3} U_i\nonumber).
  \end{eqnarray} 
From projective measurements in the computational basis, we can construct the estimator 
 \begin{align}
 \mathcal{E}_\mathrm{KB} &= 2^{2n} \sum_{\mathbf{s}_{I},\mathbf{s}_{I}'}  (-2)^{- D[\mathbf{s}_{\{I_1,I_3\}},{s}_{\{I_1,I_3\}}')]}
 \sigma_{I_2}^z(s_{I_2}) \sigma_{I_2}^z(s'_{I_2})
  \nonumber \\
 & \hspace{1cm} \overline{P_{U^{(1)}_\mathrm{KB}}(s_I)P_{U^{(2)}_\mathrm{KB}}(s'_I)},
  \nonumber \\
  &= \tr{\overline{ (U_\mathrm{KB}^{(1)})^\dag \otimes  (U_\mathrm{KB}^{(2)})^\dag\,   O_\mathrm{KB} \,U_\mathrm{KB}^{(1)}\otimes U_\mathrm{KB}^{(2)}} \,(\rho_I  \otimes \rho_I)}\nonumber \\
    &= \textrm{Tr}\left[  \bigotimes_{i\in I_1} \overline{ U_i^\dag \otimes (U_i^*)^\dag\,   O_{\mathrm{KB},i} \,U_i\otimes U_i^*} 
    \bigotimes_{i\in I_2} O_{\mathrm{KB},i}
    \right. \nonumber \\
    &\qquad \quad \left. \bigotimes_{i\in I_3} \overline{ U_i^\dag \otimes U_i^\dag\,   O_{\mathrm{KB},i} \,U_i\otimes U_i} \; (\bar{\rho}_I  \otimes \rho_I) \right],
 \end{align}
with $
    \bar{\rho}_I\equiv (u_\mathcal{T} \otimes \mathbb{1}_{I_2} \otimes \mathbb{1}_{I_3})\,\rho_I\, ( u^\dagger_\mathcal{T} \otimes \mathbb{1}_{I_2}\otimes \mathbb{1}_{I_3})$. Here,  we defined
 the operator $O_\mathrm{KB}=\bigotimes_{i\in I}O_{\mathrm{KB},i}$, with 
\begin{align*}
O_{\mathrm{KB},i\in I_1 \cup I_3} &= 2 \sum_{s_i,s_i'} (-2)^{-D[s_{i},s_{i'}]}
 \ket{s_i}\bra{s_i} \otimes  \ket{s'_i}\bra{s'_i}. \\
O_{\mathrm{KB},i\in I_2}&=\sigma_i^z \otimes \sigma_i^z.
\end{align*}
Using the random unitary twirling channel introduced in the Methods, this leads to
 \begin{align}
 \mathcal{E}_\mathrm{KB} &= \tr{ \bigotimes_{i\in I_1} \mathbb{T}_{i} \bigotimes_{i\in I_2} O_{\mathrm{KB},i} \bigotimes_{i\in I_3} \mathbb{S}_{i} (\bar{\rho}_I  \otimes \rho_I)}
 \nonumber \\ 
 &= \mathcal{Z}_\mathrm{KB}.
 \end{align}
 
 \section{Statistical errors}
  \begin{figure}[t]
    \centering
    \includegraphics[width=0.49\textwidth]{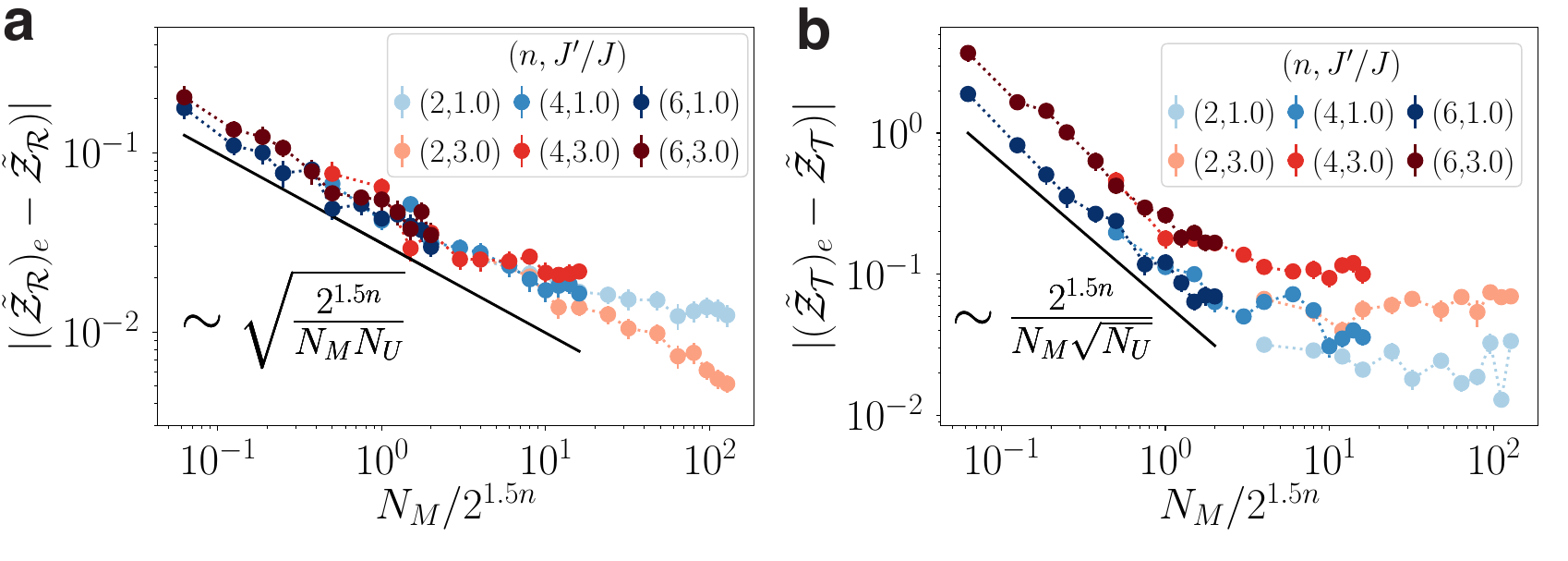}
    \caption{\textbf{Statistical Errors}
    {\bf a}, Statistical errors of $\zrn$ for $N_U=256$, and as a function of $N_M$, for different $n=4,6$ ($N=24$), and two values of $J'/J=1,3$. 
    {\bf b}, Same as {\bf a} for $\ztn$.
    The errors are obtained by averaging individual statistical errors over $32$ simulated experiments.
    }
    \label{fig:stat}
\end{figure}
In this section, we study the role of statistical errors arising in our protocols from a finite number of random unitaries $N_U$ used to estimate the ensemble average and a finite number of measurements $N_M$ per random unitary performed to estimate occupation probabilities.
We consider  the measurement of MBTIs associated with reflection and time-reversal symmetry, and numerically investigate the average statistical error of the estimation of $(\zrn)_e$, and $(\ztn)_e$.

We show in Fig.~\ref{fig:stat} the errors obtained for different values of $n$ and $J'/J$, as a function of the number of projective measurement per unitary $N_M$, for a fixed value of $N_U=250$.
We observe that the scaling $2^{1.5n}/(N_M \sqrt{N_U})$) of statistical  errors of the time reversal invariant  measured in an interval $I$ with $2n$ sites  is very similar to the scaling to access the R\'enyi entropy of $2n$ qubits which was studied in details in Ref.~\cite{Elben2018,Elben2018a}. Furthermore, the statistical error of the partial reflection invariant, scaling as $\sqrt{2^{1.5n}/(N_M N_U)}$, is generically smaller than for the time reversal invariant, since it is only a linear function in the measured probabilities.

Despite an exponential scaling of the required total number of measurements $N_M N_U$ with the partition size $2n$, 
our protocol
can be applied in current experiments of Rydberg atoms, trapped ions, or superconducting qubits offering high repetition rates (see in particular Ref.~\cite{Brydges2019}). In particular,  the protocol scales much more favourably than tomography. For example, the required value of $n\sim 4$ to achieve convergence of the MBTIs away from quantum critical points to their quantized value requires a reasonable total number of measurements $N_M N_U \approx 10^4$ to achieve an average statistical error of $\le 0.1$. 
It is also important to note that, in all cases, statistical errors can be estimated from a single realization of the experimental protocol using resampling techniques~\cite{Brydges2019}, i.e., it is not needed to study the statistics of different estimations of MBTIs, based on several repetitions of the protocols, to obtain statistical error bars.
 
\section{Probing the breaking of symmetries}
\begin{figure}
    \centering
    \includegraphics[width=0.49\textwidth]{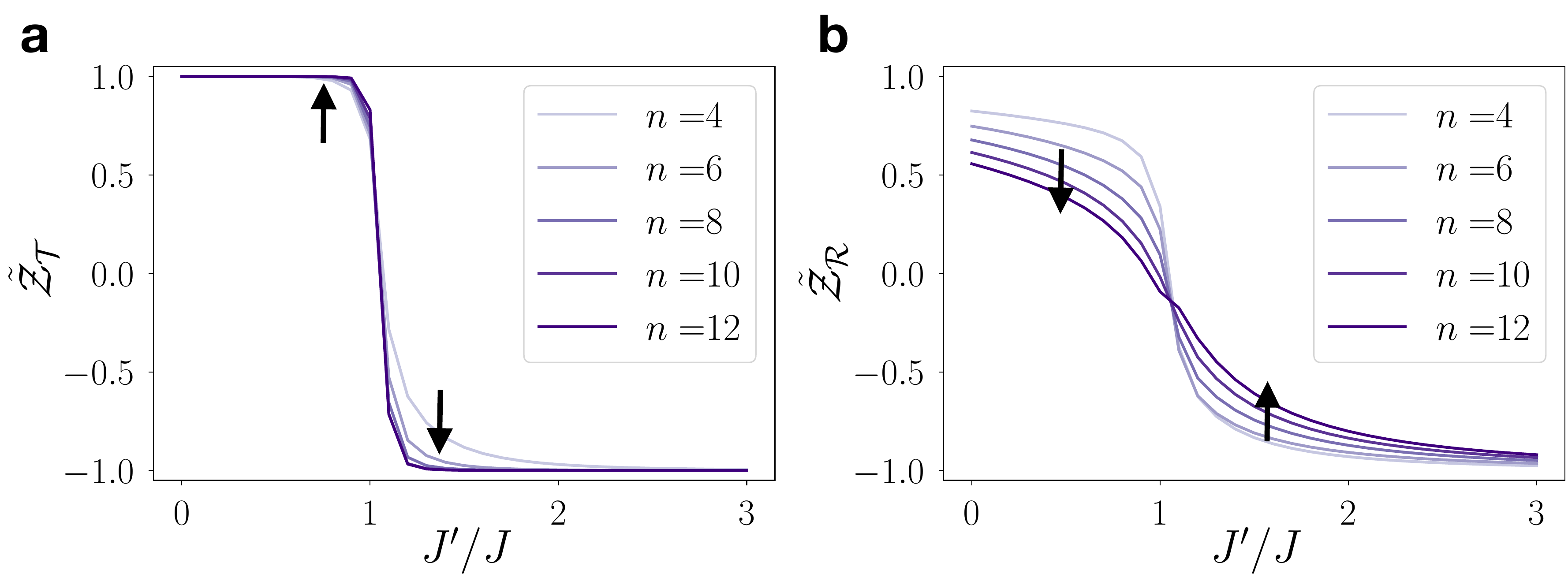}
    \caption{\textbf{Detecting the protecting symmetries for the SPT states.} In the presence of the symmetry-breaking perturbation $H_B$ [Eq.~\eqref{Eq:H_B}], the topological phase in the modified bond-alternating XXZ-model is (only) protected by the time reversal symmetry. 
    \textbf{a}, This is detected by the partial time-reversal MBTI $\tilde{\mathcal{Z}}_\mathcal{T}$ -- converging to the quantized values  $\pm1$ for increasing $n$  -- which still identifies the topological phase transition. 
    \textbf{b}, On the contrary, the partial reflection MBTI $\tilde{\mathcal{Z}}_\mathcal{R}$ -- approaching $0$ with increasing $n$ -- shows that the reflection symmetry is explicitly broken for a nonzero $B$ in Eq.~\eqref{Eq:H_B}. We choose  $B=0.1J$, $\delta=0.3$, and $N=48$. }
    \label{fig:SymBreak}
\end{figure}
Measuring MBTIs provides us with the ability to study the protection mechanism of SPT phases. 
In particular, SPT order can still exist in the absence of certain internal symmetries (thus string-order being absent), provided  at least one protecting symmetry is present~\cite{Pollman-Protecting-Symmetries2012}. 
In order to illustrate this effect with MBTIs, we add here the term 
\begin{equation} \label{Eq:H_B}
    {H_B} = B\sum_{j=1}^{N-1} {\left( {\sigma _j^x\sigma _{j + 1}^z - \sigma _j^z\sigma _{j + 1}^x} \right)}
\end{equation}
to the original Hamiltonian $H_\text{XXZ} $. In the Hamiltonian $H=H_\text{XXZ} + H_B$, the reflection and $D_2$ symmetries are explicitly broken, but the time-reversal symmetry is respected~\cite{Cooper2018}.
Thus, the ground state of $H=H_\text{XXZ} + H_B$ can still exhibit  non-trivial SPT order, protected solely by the time-reversal symmetry. This is  encoded in the values of the MBTIs, and can be thus revealed experimentally via our protocols. 
This is displayed in Fig.~\ref{fig:SymBreak},  which shows that the partial time-reversal MBTI $\ztn$ converges to $\pm 1$ for $n\to \infty$, whereas the partial reflection MBTI $\zrn$ approaches $0$ as $n\to \infty$.

\section{Adiabatic state preparation}
\begin{figure}
    \centering
    \includegraphics[width=0.49\textwidth]{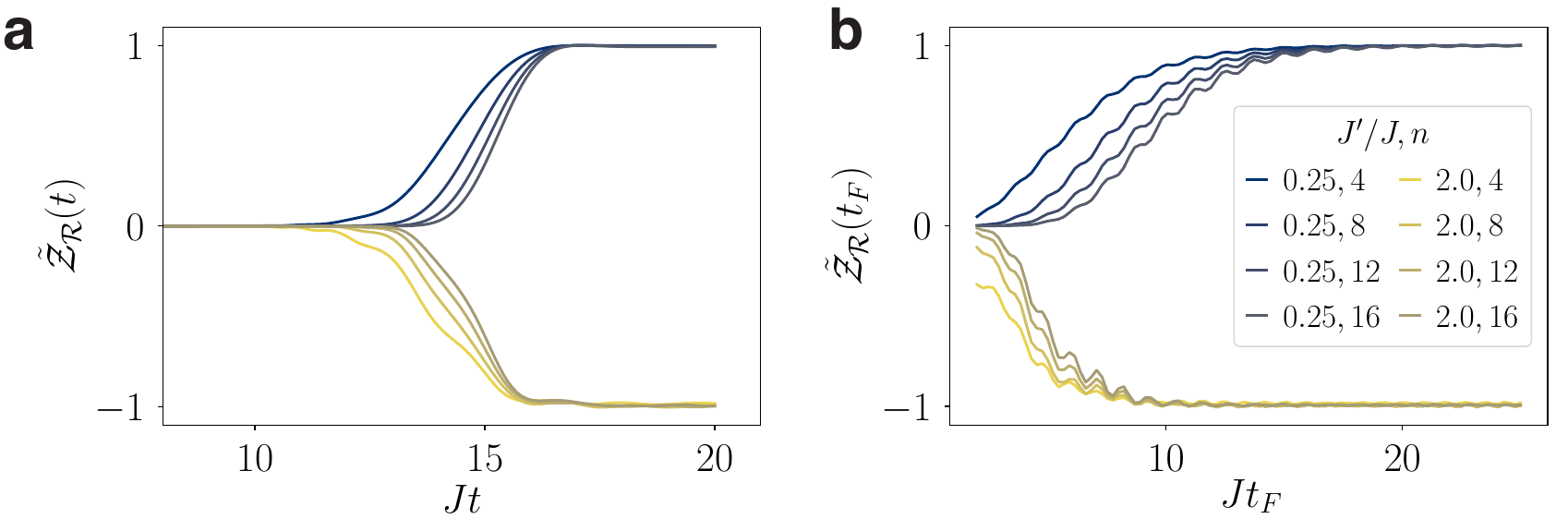}
    \caption{\textbf{Monitoring the adiabatic state preparation of an SPT state.}
    \textbf{a}, Starting from a trivial N\'eel state without reflection symmetry $\tilde{\mathcal{Z}}_\mathcal{R}(t)$, the ground state of $H_\text{XXZ}$ is adiabatically prepared (see text for details of the adiabatic ramp). This is monitored by the evolution of $\tilde{\mathcal{Z}}_\mathcal{R}(t)$ which evolves to quantized values $\pm 1$ at late times. The dynamical build-up of long-range SPT order -- for intermediate times up to a certain length scale -- is indicated at intermediate times by the increasing magnitude of $\tilde{\mathcal{Z}}_\mathcal{R}(t)$ for decreasing number $n$ of reflected pairs of spins. Here, $Jt_F=20$.
    \textbf{b}, The convergence of $\tilde{\mathcal{Z}}_\mathcal{R}(t_F)$ to $\pm 1$ as a function of the total preparation time $t_F$ indicates that for sufficiently long preparation times the ground state in trivial and topological state is prepared with high fidelity.
    For the simulations, we used the time evolving block decimation  (TEBD) algorithm. Parameters for both panels: $\delta=0.25$, $\Delta=40J$, $N=48$.
    }
    \label{fig:Preparation}
\end{figure}
In this section, we show how to prepare the ground state of our model via adiabatic state preparation, and the behavior of MBTIs during this preparation.
Here, we consider a time-dependent Hamiltonian
\begin{equation}
    H(t)=H_\mathrm{XXZ}+f(t) H_\text{N\'eel},
\end{equation}
where $H_\text{N\'eel}=\Delta \sum_i (-1)^i \sigma_i^z$, and $\Delta\gg J',J$.
At time $t=0$, the system is initialized in the N\'eel state $\ket{\psi(t=0)}=\ket{\downarrow\uparrow\downarrow\dots}$ and we set $f(0)=1$. We use the function $f(t)=(t/t_F-1)^4$ to adiabatically drive the system to the ground state of $H_\mathrm{XXZ}$ at the final time $t=t_F$. Our protocols give access to the time-dependent values of MBTI $\ztn(t), \zrn(t)$, obtained using the experimental recipe described above with random unitaries applied on the state $\ket{\Psi(t)}$.
We illustrate the emergence of quantized values of the MBTI, associated with the preparation of the SPT phases, in  Fig.~\ref{fig:Preparation}\textbf{a}. Note that the time of the preparation $Jt_F=20$ is compatible with coherence time achieved in SPT Rydberg experiments~\cite{DeLeseleuc2018}.
As shown in Fig.~\ref{fig:Preparation}\textbf{b}, the values of $\zrn(t_F)$ at the end of the preparation $t=t_F$ can be used to detect the quality of the preparation of an SPT phase:
For $Jt_F\gg 1$, the preparation is perfectly adiabatic and the values of the MBTI correspond to the ones of the ground state wave function (as presented in Figs.~\ref{fig:Figure1} and \ref{fig:Figure2} of the main text).
For $Jt_F\sim 1$, the correlations in the wave function do not extend over the full system, as in the true SPT ground-state, but only extend to certain characteristic length scale $n_c$.
Consequently, for $n\gg n_c$, the many-body invariant tends to zero. 
We expect a similar behavior for a scenario where $\ket{\Psi(t_F)}$ is replaced by a thermal state, and $n_c$ by a ``thermal length'' describing the range of correlations.

\section{Details on the DMRG simulations}
Density-Matrix-Renormalization-Group (DMRG) simulations were realized using the ITensor Library (http://itensor.org). The model was numerically solved with open-boundary conditions, with an additional small pinning field acting on the first site $\delta_p \sigma_1^z$, with $\delta_p=0.05J$, to select one of the two degenerate ground states present in the topological phase for open boundary conditions~\cite{DeChiara2012}. Note that in experiment with large system size $N$, the system would always choose one of the degenerate ground states because a cat state (i.e., the superposition of the two degenerate ground states) is always fragile to pertubations (as simulated by the small pinning field). 
We used a maximum bond dimension of $D=512$.
The exact MBTIs were extracted from direct contractions of the Matrix-Product-States representing the ground state (as shown in Figs.~\ref{fig:Figure1} and \ref{fig:Figure2} of the main text). The estimations for random measurements were obtained using a sampling algorithm of the occupation probabilities $P_U(\mathbf{s})$ for Matrix-Product-States~\cite{Han2018}.

The time-dependent simulations were realized via the Time-Evolving-Block-Decimation (TEBD) algorithm with a time step $Jdt=0.001$ and a maximum bond dimension $D=512$.
\end{document}